\documentclass[aps,prb,twocolumn,superscriptaddress,floatfix]{revtex4-1}
\usepackage{graphicx,graphics}
\usepackage{dcolumn}
\usepackage{amsmath,amssymb,amsfonts}
\usepackage{latexsym,verbatim}
\usepackage{bm}
\usepackage{color}
\usepackage{ulem}
\usepackage[percent]{overpic}
\usepackage[breaklinks=true,colorlinks,citecolor=blue,linkcolor=blue,urlcolor=blue]{hyperref}

\usepackage{array}

\begin{document}

\title{Failure of conductance quantization in two-dimensional topological insulators due to non-magnetic impurities}

\author{Pietro Novelli}
\email{pietro.novelli@sns.it}
\affiliation{Istituto Italiano di Tecnologia, Graphene Labs, Via Morego 30, I-16163 Genova,~Italy}
\affiliation{NEST, Scuola Normale Superiore, I-56126 Pisa,~Italy}

\author{Fabio Taddei}
\affiliation{NEST, Istituto Nanoscienze-CNR and Scuola Normale Superiore, I-56126 Pisa,~Italy}

\author{Andre K. Geim}
\affiliation{School of Physics \& Astronomy, University of Manchester, Oxford Road, Manchester M13 9PL, United Kingdom}

\author{Marco Polini}
\affiliation{Istituto Italiano di Tecnologia, Graphene Labs, Via Morego 30, I-16163 Genova,~Italy}
\affiliation{School of Physics \& Astronomy, University of Manchester, Oxford Road, Manchester M13 9PL, United Kingdom}

\begin{abstract}
    Despite topological protection and the absence of magnetic impurities, two-dimensional topological insulators display quantized conductance only in surprisingly short channels, which can be as short as $100~{\rm nm}$ for atomically-thin materials. We show that the combined action of short-range non-magnetic impurities located near the edges and onsite electron-electron interactions effectively creates non-collinear magnetic scatterers, and, hence, results in strong back-scattering. The mechanism causes deviations from quantization even at  zero temperature and for a modest strength of electron-electron interactions. Our theory provides a straightforward conceptual framework to explain experimental results, especially those in atomically-thin crystals, plagued with short-range edge disorder.
    \end{abstract}
    \maketitle
    
    {\it Introduction.---}Research on spin-orbit coupling in graphene led Kane and Mele~\cite{kane_prl_2005_146802,kane_prl_2005_226801} to predict the existence of two-dimensional (2D) topological insulators (TIs).
    These are electron systems with a gap in the bulk density of states (DOS) and pairs of conducting edge states displaying helicity, i.e.~spin-momentum locking.
    Because of Kramers theorem, in the absence of many-particle effects non-magnetic impurities in a 2DTI cannot induce back-scattering at a 2DTI edge, yielding conductance quantization against elastic disorder~\cite{xu_prb_2006,bernevig_science_2006,hasan_rmp_2010,qi_rmp_2011,bernevig_book_2013,ren_rpp_2016}.
    
    All experimental measurements on 2DTIs, however, show deviations from the expected quantized value of conductance $2e^{2}/h$, particularly in small-gap semiconductor heterostructures such as HgTe/CdHgTe and InAs/GaSb quantum wells~\cite{konig_science_2007,roth_science_2009, suzuki_prb_2013, grabecki_prb_2013,du_prl_2015}, but also in atomically-thin crystals such as ${\rm WTe}_2$~\cite{wu_science_2018, shi_arxiv_2018}.
    On the other hand, the existence of conducting edge modes was clearly demonstrated via non-local measurements in Refs.~\onlinecite{roth_science_2009,suzuki_prb_2013,grabecki_prb_2013,du_prl_2015}.
    Semiconducting heterostructures were extensively studied in the low-temperature regime (below $4~{\rm K}$)~\cite{konig_science_2007,du_prl_2015} because of their small energy gap.
    For channel lengths $L$ shorter than $\sim 1~{\rm \mu m}$, fluctuations of the conductance around the quantized value $2e^2/h$ were observed as a function of the back gate voltage.
    For longer channels, even the average conductance was found to deviate from $2e^{2}/h$ and even totally suppressed~\cite{konig_prx_2013}, when the edge was perturbed by a scanning tip.
    Among the 2DTIs realized by semiconducting heterostructures, the best results were obtained thanks to $\rm{Si}$ doping~\cite{du_prl_2015}.
    In these samples, conductance is quantized up to 1-2\% at very low temperatures. Monolayers of ${\rm WTe}_{2}$  exhibit~\cite{wu_science_2018} conductance quantization up to $100~{\rm K}$, making them the 2DTIs existing at the highest temperatures up to date, though displaying quantization only in short channels ($L \lesssim 100~{\rm nm}$).

    \begin{figure}[t]
    \centering
    \includegraphics[width=0.40\textwidth]{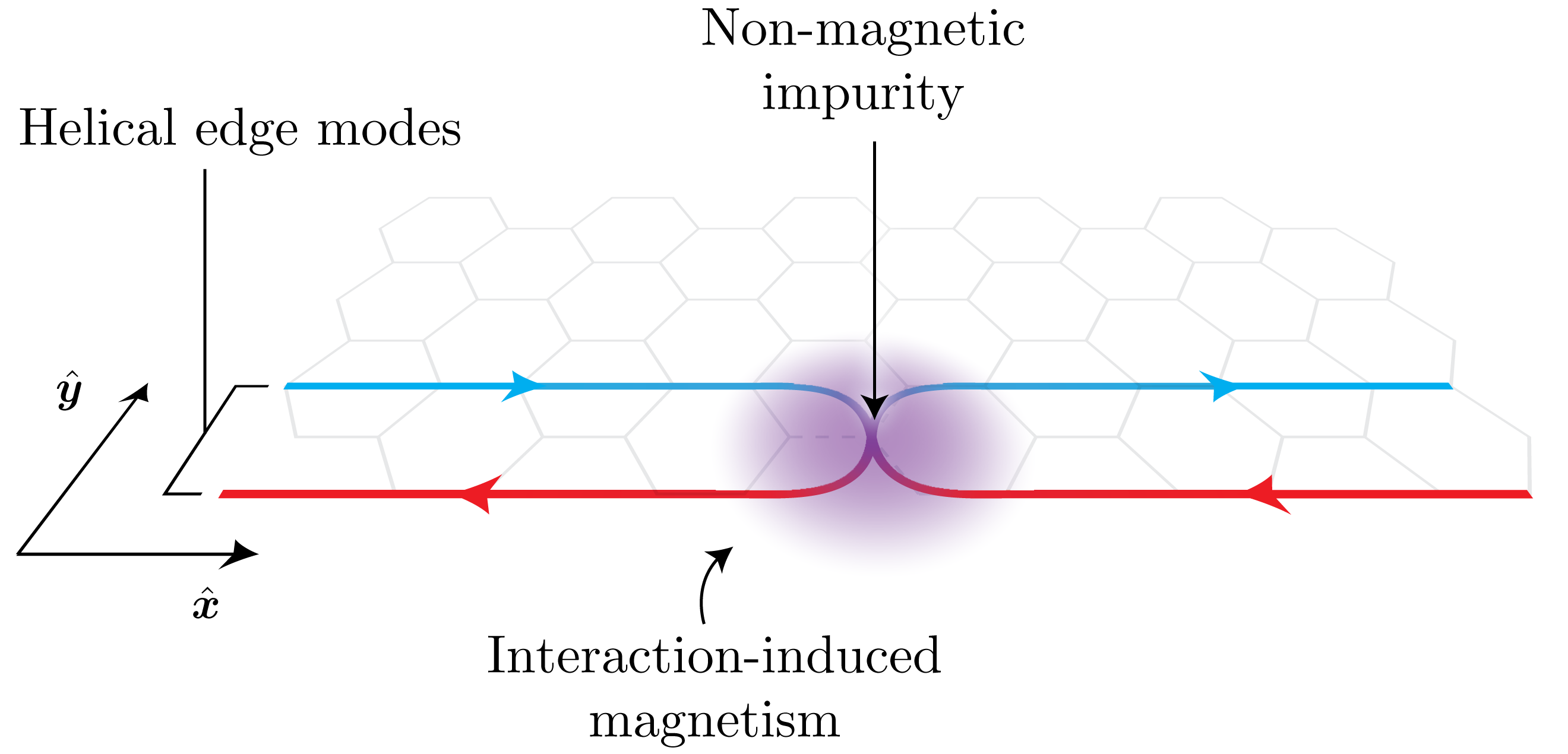}
    \caption{(Color online) A cartoon of the physical process introduced and analyzed in this work. 
    At an edge of a 2DTI, a 
    non-magnetic short-range impurity can effectively act as a magnetic one due to its dressing via onsite electron-electron interactions. 
    The latter favor the formation of a local magnetic moment with non-zero in-plane components. 
    These cause spin mixing and hence back-scattering.\label{fig:device}}
    \end{figure}

    The cause of the breakdown of conductance quantization is still poorly understood.
    Clearly, one possibility is the presence of an external magnetic field \cite{konig_science_2007,wu_science_2018} or of magnetic impurities~\cite{tanaka_prl_2011,maciejko_prl_2009,altshuler_prl_2013}, which induce spin-flip scattering (thus back-scattering).
    Magnetic impurities, however, are rare both in materials grown by molecular beam epitaxy~\cite{konig_science_2007,roth_science_2009,suzuki_prb_2013, grabecki_prb_2013,du_prl_2015} and in mechanically-exfoliated crystals~\cite{wu_science_2018, shi_arxiv_2018}, but explain experimental data in the ``extrinsic'' case in which magnetic dopants are deliberately added to pristine three-dimensional TI samples~\cite{chen_science_2010,wray_natphys_2011}.
    Coupling between opposite edges, in very narrow samples or in purposely fabricated point contacts, can also induce back-scattering~\cite{strom_prl_2009,schmidt_prl_2011,hou_prl_2009}, with no need of time-reversal symmetry breaking.
    Importantly, the breakdown of conductance quantization could arise from two-body interactions.
    In Ref.~\onlinecite{xu_prb_2006} it was suggested that electron-electron (e-e) interactions in 2DTIs can cause back-scattering through a third-order perturbation-theory scattering process, while the spontaneous breaking of time-reversal symmetry due to interactions was studied in Ref.~\onlinecite{wu_prl_2006}. Interactions are also at the core of other mechanisms proposed to explain the spoiling of conductance quantization in 2DTIs.
    Back-scattering resulting from weak e-e interactions and an impurity potential, in the absence of axial spin symmetry, was considered in Ref.~\onlinecite{schmidt_plr_2012}. Deviations from $2e^2/h$ were found to scale like $T^{4}$, at low temperatures $T$.
    The coupling of edge modes to charge puddles, naturally present in real samples, was accounted for in Refs.~\onlinecite{vayryen_prl_2013, vayryen_prb_2014} and found to lead to a correction to the conductance scaling like $T^{4}$ at low temperatures.
    In contrast, recent experiments~\cite{wu_science_2018} show nearly temperature-independent conductance in 2DTIs.
    Another mechanism that leads to the breakdown of conductance quantization is related to the edge reconstruction~\cite{wang_prl_2017}, which can occur when the confining potential of the 2DTIs edges is not sufficiently sharp.  Finally, the effects of Rashba spin-orbit coupling~\cite{strom_prl_2010,crepin_prb_2012}, phonons~\cite{budich_prl_2012}, nuclear spins~\cite{hsu_prb_2017,hsu_prb_2018}, disordered probes~\cite{mani_prapplied_2016}, coupling to external baths~\cite{bagrov_2018}, and noise~\cite{vayryen_2018} have also been analyzed.
    
    In this Letter we propose a simple mechanism, based on the interplay between non-magnetic scatterers and e-e interactions, which leads to the breakdown of conductance quantization in 2DTIs, even at zero temperature, and can result in the total suppression of the conductance.
    Starting from the single-particle Kane-Mele Hamiltonian~\cite{kane_prl_2005_146802,kane_prl_2005_226801} describing a 2DTI ribbon, we consider the presence of short-range non-magnetic impurities at its edges (see Fig.~\ref{fig:device}). As expected, this leads to an enhancement of the local DOS, as in the case of midgap states in graphene~\cite{katsnelson} and three-dimensional TIs~\cite{black-schaffer_prb_2014,black-schaffer_prb_2012,black-schaffer_prb_2012a}.
    In the presence of Hubbard-like e-e interactions, using the self-consistent unrestricted Hartree-Fock method, we show that these short-range defects favor the formation of local magnetic moments, leading to the spontaneous breakdown of time-reversal symmetry and back-scattering.

    {\it Theoretical model.---}We consider the Kane-Mele-Hubbard model~\cite{rachel_prb_2010,rachel_arXiv_2018},
    \begin{equation}\label{eq:full_hamiltonian}
        \begin{split}
            \mathcal{H} = & ~ t\sum_{\langle ij \rangle,  \alpha}c^{\dagger}_{i \alpha}c_{j \alpha} + i\lambda \sum_{\langle \langle ij \rangle \rangle, \alpha, \beta}\nu_{ij} c^{\dagger}_{i\alpha}\sigma^{z}_{\alpha\beta}c_{j\beta}\\ & + ~ U\sum_{i}c_{i\uparrow}^{\dagger}c_{i\uparrow}c_{i\downarrow}^{\dagger}c_{i\downarrow}~.
        \end{split}
    \end{equation}
    In Eq.~(\ref{eq:full_hamiltonian}), $c_{i\alpha}^{\dagger}~(c_{i\alpha})$ creates (destroys) an electron of spin $\alpha$ on the $i$-th site of a honeycomb lattice and $\sigma^{z}$ is a $2 \times 2$ Pauli matrix acting on spin space.
    The sums over $\langle ij\rangle~(\langle \langle ij\rangle\rangle)$  are intended between $i$ and $j$ being first (second) neighbours. The parameters $t$ and $\lambda$ are hopping energies between first and second neighboring sites, respectively.
    The second term in Eq.~(\ref{eq:full_hamiltonian}) was introduced by Kane and Mele~\cite{kane_prl_2005_146802,kane_prl_2005_226801} as a time-reversal invariant version of the Haldane model~\cite{haldane_prl_1988}, and is responsible for the existence of helical edge modes. The factor $\nu_{ij}$ is equal to $\pm 1$, with $\nu_{ji}=-\nu_{ij}$, depending on the orientation of the two nearest-neighbor bonds the
    electron traverses in going from site $j$ to $i$: $\nu_{ij}=-1(+1)$ if the electron reaches the second neighbour going (anti-)clockwise.
    The last term accounts for local e-e repulsive interactions. Such a two-body term will be treated within mean-field theory. The key point here is that we are not interested in dealing accurately with strong correlations in 2DTIs~\cite{rachel_arXiv_2018}. Our aim is to utilize the simplest approach that enables us to capture an important effect stemming from local e-e interactions in the weak-coupling $U/t< 1$ regime. In this regime, mean-field theory is expected to be accurate and allows us to obtain an effective single-particle Hamiltonian, which can be used in combination with Landauer-B\"uttiker theory~\cite{datta_book_1995} to compute transport properties. 
    
We consider a ribbon extending in the region $0 \leq x \leq L$, $0 \leq y \leq W$, with armchair edges and periodic boundary conditions in the $\hat{\bm x}$-direction (see Fig.~\ref{fig:device}). 
In order to investigate the effect of atomic-scale defects, we assume the presence of one or two vacancies, which can be accounted for by dropping from the sums in Eq.~(\ref{eq:full_hamiltonian}) terms involving the lattice sites where the atoms are missing. The case of many vacancies can be tackled in a straightforward manner but lies beyond the scope of this work.  Our main point, here, is to demonstrate the importance of local e-e interactions in dressing short-range non-magnetic impurities in a magnetic fashion.
    
 Using the usual Hartree-Fock decoupling~\cite{Giuliani_and_Vignale}, we can express (\ref{eq:full_hamiltonian}) in the unrestricted Hartree-Fock approximation~\cite{verges_prb_1991, verges_prb_1992} as
    \begin{equation}\label{eq:main_hamiltonian}
        \begin{split}
            \mathcal{H} \simeq & ~ t\sum_{\langle ij \rangle,  \alpha}c^{\dagger}_{i \alpha}c_{j \alpha} + i\lambda \sum_{\langle \langle ij \rangle \rangle, \alpha, \beta}\nu_{ij}c^{\dagger}_{i\alpha}\sigma^{z}_{\alpha\beta}c_{j\beta}\\ & + ~ \frac{U}{2}\sum_{i,\alpha, \beta}c^{\dagger}_{i\alpha}(n_{i}\openone_{\alpha\beta}- \boldsymbol{m}_{i}\cdot {\bm \sigma}_{\alpha\beta})c_{i\beta} \\
            &- \frac{U}{4}\sum_{i}(n^2_{i} - |\boldsymbol{m}_{i}|^2)~,
        \end{split}
    \end{equation}
    where $\openone$ is the $2 \times 2$ identity matrix, $\boldsymbol{\sigma}=(\sigma^{x}, \sigma^{y}, \sigma^{z})$ is a vector of $2\times 2$ Pauli matrices acting on spin space, and we have defined the local mean electron density
    \begin{equation}\label{eq:gs_density}
    n_{i} = \langle \sum_{\alpha}c^{\dagger}_{i\alpha}c_{i\alpha} \rangle
    \end{equation}
    and the local mean spin polarization ${\bm s}_i = \hbar \boldsymbol{m}_{i}/2 = \hbar (m^{x}_{i},m^{y}_{i},m^{z}_{i})/2$ with
    \begin{equation}\label{eq:gs_magnetization}
    \boldsymbol{m}_{i} = \langle \sum_{\alpha, \beta}c^{\dagger}_{i\alpha}{\bm \sigma}_{\alpha\beta}c_{i\beta}\rangle~,
    \end{equation}
    which must be determined self-consistently. In order to do so, we use an iterative algorithm~\cite{feldner_prb_2010} which involves the exact diagonalization of the Hamiltonian (\ref{eq:main_hamiltonian}).
    Our calculations were performed at $T = 0$, but can easily be extended to finite temperature. Technicalities are reported in Appendix \ref{sec:technical_details}.
    For $\lambda = 0$, i.e.~when the second neighbour hopping term is neglected, the lattice is bipartite in the sense of Ref.~\onlinecite{lieb_prl_1989} and Lieb theorem holds, so that a non-zero ground-state spin polarization rigorously follows from sublattice imbalance (i.e.~different number of sites in the two sublattices).
    As we will see below, a ground-state spin polarization occurs even for $\lambda \neq 0$---i.e.~in the topological phase of (\ref{eq:full_hamiltonian}) with gap~\cite{kane_prl_2005_146802} $\delta_{\rm g} = |6\sqrt{3}\lambda|$---where Lieb theorem does not apply. All numerical results below refer to a rectangular sample with $L=45(\sqrt{3}/2)a$ and width $W=25a$.
    
\begin{figure}[t]
    \centering
    \includegraphics[width=0.49\textwidth]{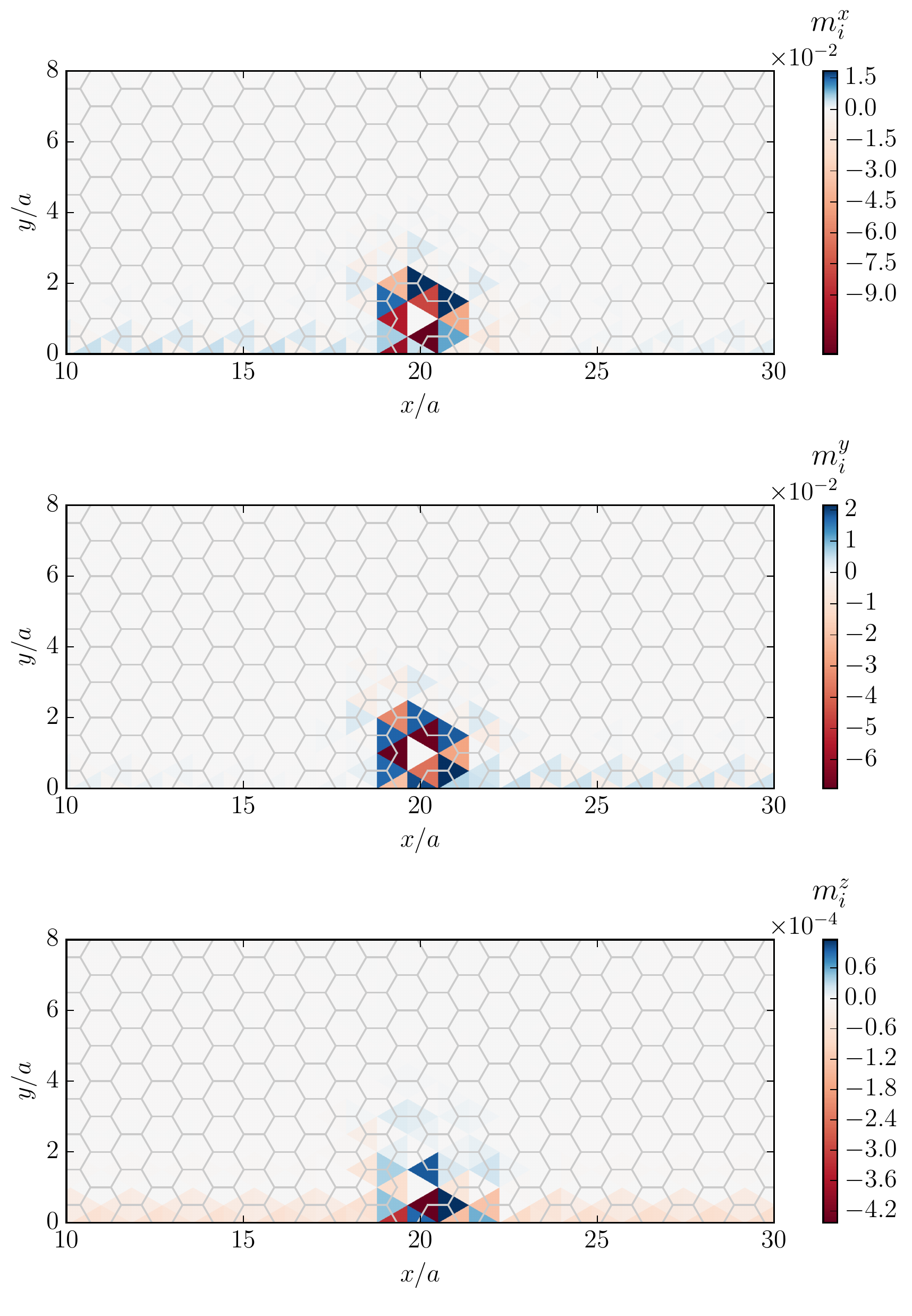}
    \caption{(Color online) Interaction-induced spin polarization near a vacancy. Color plot of the three components of the spatial profile of the dimensionless spin polarization ${\bm m}_i$ around a vacancy located at $x=23 (\sqrt{3}/2)a$ and $y=a$. Top panel: $m^{x}_i$. Central panel: $m^{y}_i$. Bottom panel: $m^{z}_i$. From Eq.~(\ref{eq:main_hamiltonian}) it is clear that the components of ${\bm m}_{i}$ lying on the $\hat{\bm x}$-$\hat{\bm y}$ plane are those leading to spin-mixing and hence back-scattering. Numerical results in this figure have been obtained by setting $\lambda/t = 0.09$ and $U/t = 0.1$.\label{fig:magnetization}}
    \end{figure}

    {\it Ground-state spin polarization.---}In Fig.~\ref{fig:magnetization} we plot the spatial profile of the three components---$m^{x}_{i}$, top panel, $m^{y}_{i}$, central panel, and $m^{z}_{i}$, bottom panel---of the dimensionless spin polarization (\ref{eq:gs_magnetization}), calculated at half filling for $\lambda/t = 0.09$ and $U/t = 0.1$, when a single vacancy is placed at $x=23 (\sqrt{3}/2)a$ and $y=a$, where $a$ is the lattice parameter. The ground-state electron density $n_{i}$ turns out to be nearly uniform.
    
    The results show that spin polarization occurs around the vacancy, being vanishing elsewhere with the exception of asymmetric tails extending throughout the edge. This nicely agrees with the Stoner criterion, stating that a ground-state magnetization can occur in presence of a peak in the DOS. Indeed, a short-range defect generally hosts bound states localized around it, leading to an enhancement of the local DOS in proximity of the defect.

    It is interesting to note that a finite spin polarization is bound to atomic-scale imperfections. Away from the vacancy the sample displays zero spin polarization. We thus expect that short-range edge roughness, which naturally occurs e.g.~in atomically-thin crystals~\cite{wu_science_2018, shi_arxiv_2018}, as well can in general lead to interaction-induced spin polarization. We now move to analyze its effects on the transport properties of the system.
    
    {\it Breakdown of conductance quantization.---}Due to spin-momentum locking, back-scattering is induced by spin-flip events, which, in turn, are induced by the terms proportional to $m^{x}_{i}$ and $m^{y}_{i}$ in Eq.~(\ref{eq:main_hamiltonian}).
    Once the mean-field theory parameters $n_{i}$ and ${\bm m}_i$ are obtained, the conductance of the sample in a two-terminal setup (where one lead is attached to the left and the other to the right) can be calculated within the Landauer-B\"uttiker formalism~\cite{datta_book_1995}.
    In particular, at zero temperature, the differential conductance $G$ is given by $G=(2e^2/h) {\cal T}$, ${\cal T}$ being the transmission coefficient.
    Quantization of conductance is a consequence of ${\cal T}$ being an integer number.
    We have calculated ${\cal T}$ as a function of energy $E$ for the mean-field Hamiltonian (\ref{eq:main_hamiltonian})---with $n_{i}$ and $\boldsymbol{m}_{i}$ calculated self-consistently---by utilizing the toolkit ``KWANT''~\cite{groth_njp_2014}. 
    The leads are defined by the same Hamiltonian (\ref{eq:main_hamiltonian}) with ${\bm m}_{i} ={\bm 0}$ and $n_{i}$ uniform and equal to $1$ (corresponding to half filling) for every $i$. 
    
\begin{figure}[t]
    \centering
    \includegraphics[width=0.49\textwidth]{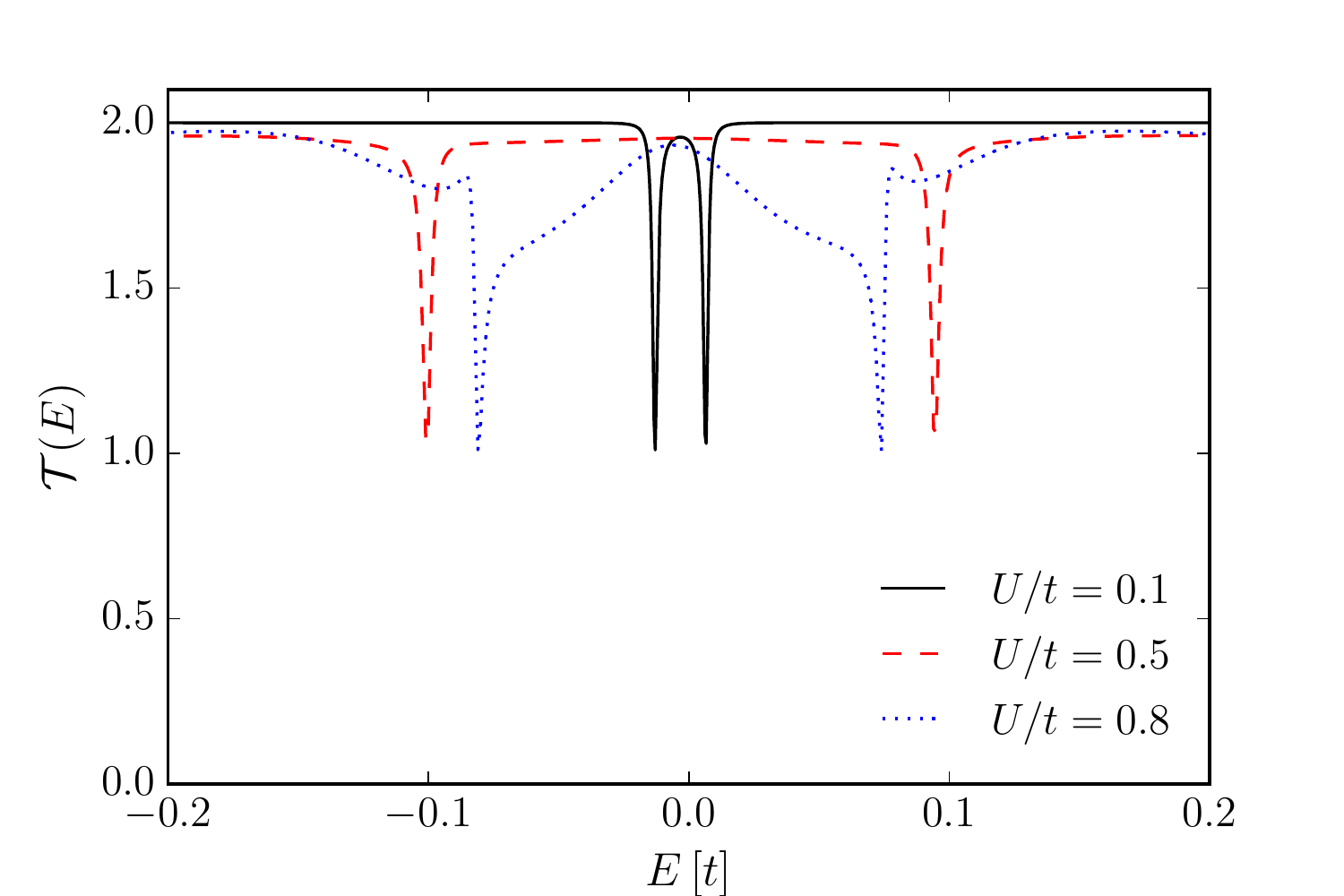}
    \caption{(Color online) Breakdown of conductance quantization for a single vacancy at the edge of a 2DTI. The transmission ${\cal T}$ is plotted as a function of energy $E$ (in units of $t$) at half filling and for energies lying in the gap $\delta_{\rm g}$. Different curves refer to different values of $U/t$. Numerical results in this figure have been obtained by setting $\lambda/t = 0.09$ ($\delta_{\rm g}\simeq 0.93 t$). Since on-site e-e interactions produce a spin polarization with in-plane components near the vacancy, back-scattering events occur at the same 2DTI edge and lead to the breakdown of conductance quantization, i.e.~${\cal T}<2$. \label{fig:single_vacancy_transmission}} 
    \end{figure}

\begin{figure}[t]
    \centering
    \includegraphics[width=0.49\textwidth]{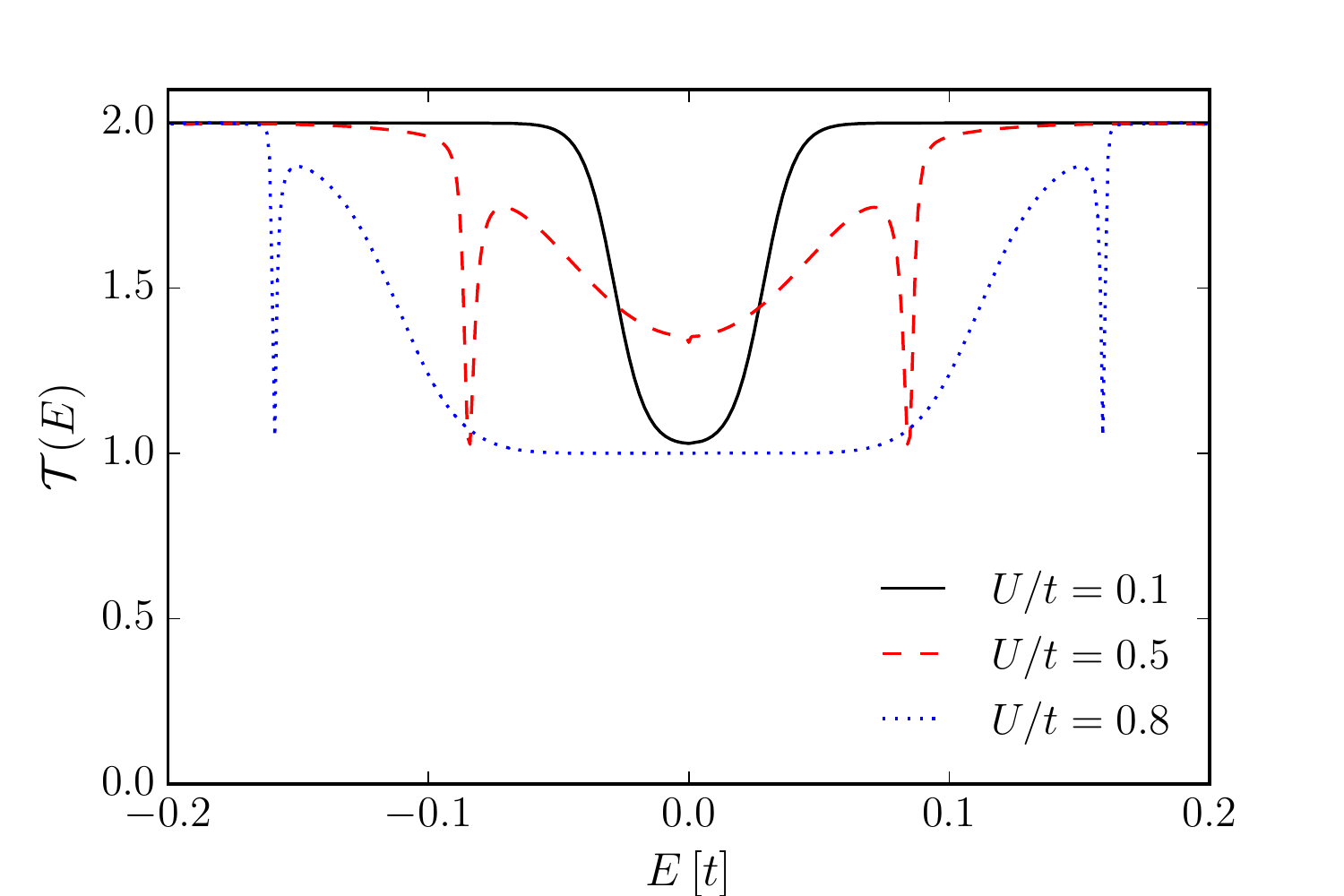}
    \caption{(Color online) Same as in Fig.~\ref{fig:single_vacancy_transmission} but for the case of two vacancies placed at $x=23 (\sqrt{3}/2)a$, $y=(3/2)a$ and $x=26 (\sqrt{3}/2)a$, $y=(1/2)a$.\label{fig:double_vacancy_transmission}} 
    \end{figure}

    Figs.~\ref{fig:single_vacancy_transmission} and~\ref{fig:double_vacancy_transmission} show the transmission coefficient ${\cal T}$ as a function of energy $E$ ($E=0$ denotes the energy at which the edge-mode dispersions cross in the leads), in the presence of one and two vacancies, respectively, and for different values of $U/t$. According to Fig.~\ref{fig:single_vacancy_transmission}, relative to a single vacancy placed at $x=23 (\sqrt{3}/2)a$ and $y=a$, ${\cal T}<2$ (thus conductance quantization is spoiled) for $E\approx 0$.
    In particular, pairs of sharp dips appear where back-scattering is maximum and ${\cal T}$ takes its minimum value, i.e.~${\cal T}\simeq 1$ due to the presence of an unperturbed propagating mode on the opposite edge of the sample.
    The main effect of increasing $U$ from $0.1t$ to $0.5t$ is an enhancement of the separation between the dips, while the value of ${\cal T}$ between the dips is slightly suppressed (by a few percent) with respect to ${\cal T}=2$, 
    virtually independently of $U$.
    For larger values of $U$, for example at $U=0.8t$, ${\cal T}$ is much more affected presenting, apart from the pairs of dips, a sensible suppression in a larger range of energies.
    A few remarks are in order here.
    First, due to the approximate particle-hole symmetry of the model (\ref{eq:full_hamiltonian}) at $\lambda/t\ll 1$, the transmission is a nearly perfectly even function of $E$.
    As already noted, the transmission is never below $1$ because the unperturbed edge mode on the opposite side of the ribbon is perfectly conducting.
    Notice that at the energies where the dips occur the transmission relative to one edge mode nearly vanishes. Nearly total suppression of the conductance in a 2DTI was experimentally observed in Ref.~\onlinecite{konig_prx_2013}.
    Since the sample displays a finite spin polarization only around the impurity and the edge-mode wavefunctions decay exponentially away from the edge, the detrimental effects of a vacancy on $G$ rapidly vanish as this is moved towards the center of the sample (see Appendix \ref{sec:impurity_distance}).

    The behavior of ${\cal T}$ for $U/t\ll 1$ can be understood by solving the problem of a magnetic $\delta$-like impurity~\cite{dang_J_Phys_Condens_Matter_2016,zheng_prb_2018} at a single edge. In this regime, the dips in ${\cal T}(E)$ can be parametrized by a Breit-Wigner dependence on $E$, as shown in Appendix \ref{sec:numerical_fit}. 
    Accordingly, such dips can be explained as anti-resonances resulting from the localization of an electron around the impurity. Local DOS calculations (see Appendix \ref{sec:ldos}) show that at the energy $E = \pm E_{\rm a}$ of the dips the local DOS peaks around the impurity. This suggests that an electron with energy $E = \pm E_{\rm a}$ traversing the sample gets localized in the bound state around the impurity and scattered back after a waiting time, which is inversely proportional to the width of the Breit-Wigner function.

    Fig.~\ref{fig:double_vacancy_transmission} shows the transmission calculated in the presence of two vacancies. We clearly see that ${\cal T}$ is much more affected by the vacancies with respect to the case of a single vacancy, being suppressed in larger ranges of energy even in the weak-coupling regime. Moreover, for $U=0.8t$ the transmission relative to one edge mode is suppressed to zero for $-0.1t<E<0.1t$.
    
    {\it Summary and discussion.---}We have shown that the combined action of short-range non-magnetic impurities and onsite e-e interactions in two-dimensional topological insulators leads to strong back-scattering. 
    
    Strong deviations from quantization occur even in the zero-temperature limit. In contrast, all other theories~\cite{wu_prl_2006,schmidt_plr_2012,vayryen_prl_2013,vayryen_prb_2014} including e-e interactions yield deviations of the conductance from its quantized value, which vanish rapidly (i.e.~like $T^{\alpha}$ with $\alpha \geq 4$) as a function of temperature $T$ in the low-temperature limit. These deviations, scaling as power-laws of $T$, arise because of scattering processes induced by e-e interactions. In the present work, on the other hand, we have shown that the ground state of the Kane-Mele-Hubbard model displays a $T=0$ quantum phase transition from a paramagnetic to a magnetic state if short-range impurities and onsite e-e interactions are taken into account. It is because of this ground-state quantum phase transition that our corrections to conductance quantization do not scale to zero in the $T\to 0$ limit. Ground-state edge reconstruction due to e-e interactions~\cite{wang_prl_2017} also operates down to $T=0$ but applies only to samples with smooth confining potentials. For example, for a BHZ model applied to a HgTe/CdHgTe quantum well~\cite{bernevig_science_2006}, edge reconstruction occurs~\cite{wang_prl_2017} for confining potentials that decay slower than $13~{\rm meV}/{\rm nm}$. While certainly relevant for samples with smooth edges, the scenario of edge reconstruction is not expected to apply to atomically-thin crystals~\cite{wu_science_2018, shi_arxiv_2018}, which possess sharp edges either created naturally by mechanical exfoliation or deliberately by etching.
    
    In our theory, large deviations from quantization occur also in the weak-coupling $U/t<1$ regime, where our mean-field theory is expected to be accurate. In this case, the suppression of transmission as a function of energy can be interpreted in terms of anti-resonances stemming from the time spent by an electron in the bound states formed near short-range impurities, before is back-scattered due to spin-flipping terms in Eq.~(\ref{eq:main_hamiltonian}). 
    
    The formation of local magnetic moments in the presence of short-range impurities and onsite e-e interactions is a general feature of bipartite lattices~\cite{lieb_prl_1989,katsnelson}, for which the spectrum is particle-hole symmetric.
    Deep in the gap, any topological insulator possesses approximate particle-hole symmetry around the energy at which the edge modes cross.
    We have shown that small deviations from exact particle-hole symmetry (e.g.~due to $\lambda\neq 0$ in our model) do not spoil the formation of local magnetic moments near short-range impurities.
    Furthermore, the same happens with the addition of Rashba spin-orbit coupling, which introduces extra terms breaking the exact particle-hole symmetry of (\ref{eq:full_hamiltonian}) at $\lambda =0$, as shown in Appendix \ref{sec:rashba_soc}.
    We therefore expect that the spontaneous formation of local magnetic moments near short-range impurities induced by onsite e-e interactions is a general feature of 2D topological insulators. In any event, recent work~\cite{marrazzo_prl_2018} has shown that a naturally occurring layered mineral (jacutingaite) realizes the Kane-Mele model.

    {\it Acknowledgements.---}We wish to thank M.I. Katsnelson and M. Gibertini for useful discussions.
    This work was supported by the SNS-WIS joint lab ``QUANTRA'' and the European Union's Horizon 2020 research and innovation programme under grant agreement No. 785219 - GrapheneCore2.

\appendix 
\section{Technical details on the self-consistent numerical algorithm}
\label{sec:technical_details}
Four mean-field parameters per site have to be determined self-consistently, namely the local mean electron density $n_{i}$ and the three Cartesian components of the local dimensionless mean spin polarization $\boldsymbol{m}_{i}$ [see Eqs.~(\ref{eq:gs_density}) and~(\ref{eq:gs_magnetization})].

The iterative algorithm that we employ to determine self-consistently such mean field parameters proceeds as follows.
We start from an initial set of values $n^{(0)}_i$ and $\boldsymbol{m}^{(0)}_{i}$ and diagonalize exactly the Hamiltonian (\ref{eq:main_hamiltonian}).
The obtained eigenstates are then used to calculate the expectation values $n^{(1)}_i$ and $\boldsymbol{m}^{(1)}_{i}$ in Eqs.~(\ref{eq:gs_density}) and~(\ref{eq:gs_magnetization}), which are then re-inserted in the Hamiltonian.
The procedure repeats until convergence is reached, i.e.~until the mean values of the $(n-1)$-th iteration coincide, up to a desired tolerance $\epsilon$, to the mean values of the $n$-th iteration.
To check convergence we use the uniform norm
\begin{equation}\label{eq:absoluteNorm}
\sup_{i}|X^{(n)}_{i} - X^{(n-1)}_{i}| \leq \epsilon~,
\end{equation}
where $X^{(n)}_{i}$ denotes the value of a mean-field parameter at the $n$-th iteration.

Numerical results reported in the main text have been obtained by setting $\epsilon=10^{-10}$.

\section{Dependence of the numerical results on the position of the impurities}
\label{sec:impurity_distance}
\begin{figure}[t]
\centering
\includegraphics[width=0.5\textwidth]{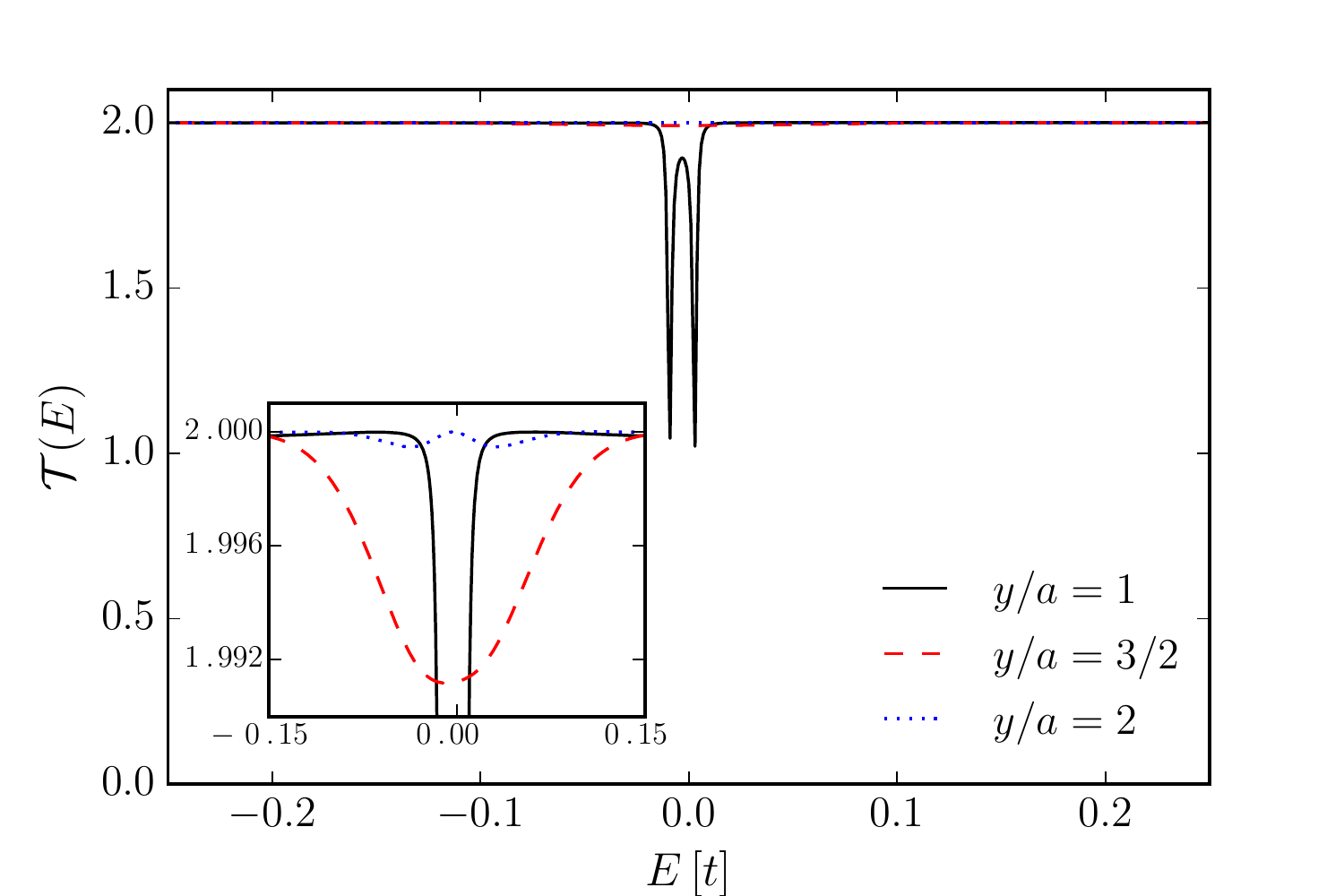}
\caption{(Color online) Transmission ${\cal T}$ as a function of energy $E$ (in units of $t$) in the presence of a single vacancy, at half filling and for energies lying in the gap $\delta_{\rm g}$. Different curves refer to different values of the distance $y$ of the vacancy from the edge, located at $y = 0$. Numerical results in this figure have been obtained by setting $\lambda/t = 0.09$ and $U/t = 0.1$. Inset: zoom of the data in the main panel.\label{fig:single_vacancy_transmission_dist}}
\end{figure}
As the impurity is moved away from the edge into the center of the sample, the suppression of the transmission ${\cal T}$ becomes negligible. This is shown in Fig.~\ref{fig:single_vacancy_transmission_dist}, where the three curves refer to three different positions of a single vacancy.
The main panel shows that when $y=(3/2)a$ and $y=2a$ the corresponding transmissions are virtually energy-independent.
A zoom of the data in the main panel is reported in the inset. For $y=2a$, the transmission deviates from $2$ by less than $0.1$\%.
This behavior is easily explained by remembering that the edge-mode wavefunctions decay exponentially away from the edge over a length scale on the order of the lattice parameter $a$ (see Fig.~\ref{subfig:1Vac_ldos_edge}).
If the distance between the edge and the vacancy is larger than $a$, their overlap decreases exponentially, strongly reducing the chances of back-scattering events.

\section{Fitting numerical data with an analytical model}
\label{sec:numerical_fit}

Our numerical data in the weak-coupling $U/t\ll 1$ regime can be explained by utilizing a simple model proposed in Ref.~\onlinecite{dang_J_Phys_Condens_Matter_2016}, which describes a single edge of a 2DTI in which a pair of helical edge modes is coupled to a $\delta$-like magnetic impurity. According to Ref.~\onlinecite{dang_J_Phys_Condens_Matter_2016}, the transmission ${\cal T}_{\rm SE}(E)$ relative to a single edge is given by
\begin{equation}\label{eq:analytical_transmission}
{\cal T}_{\rm SE}(E) = 1 - (1- \alpha^{2})\frac{\tilde{\gamma}^{2}}{(E^{2} - E_{\rm a}^{2})^{2} + \tilde{\gamma}^2}~.
\end{equation}
Here, $\pm E_{\rm a}$ with $E_{\rm a} = \sqrt{\Delta^{2} - \gamma^{2}}/2$ are the positions of the dips. The parameter $\gamma$ describes coupling between the edge mode and the magnetic impurity and $\Delta$ is the strength of the magnetic impurity, while $\tilde{\gamma} = \gamma \Delta/2$ is related to the dips' width, which in the limit $\tilde{\gamma}/E^{2}_{\rm a} \ll 1$ is given by $\tilde{\gamma}/E_{\rm a}$. The quantity $\alpha$, which take values in the range $-1 \leq \alpha \leq 1$, controls the depth of the dips and is given by $\alpha = \cos(\theta)$, where $\theta$ is the angle formed by the magnetization of the impurity with a vector normal to the plane of the 2DTI.

\begin{figure}[t]
\centering
\includegraphics[width=0.5\textwidth]{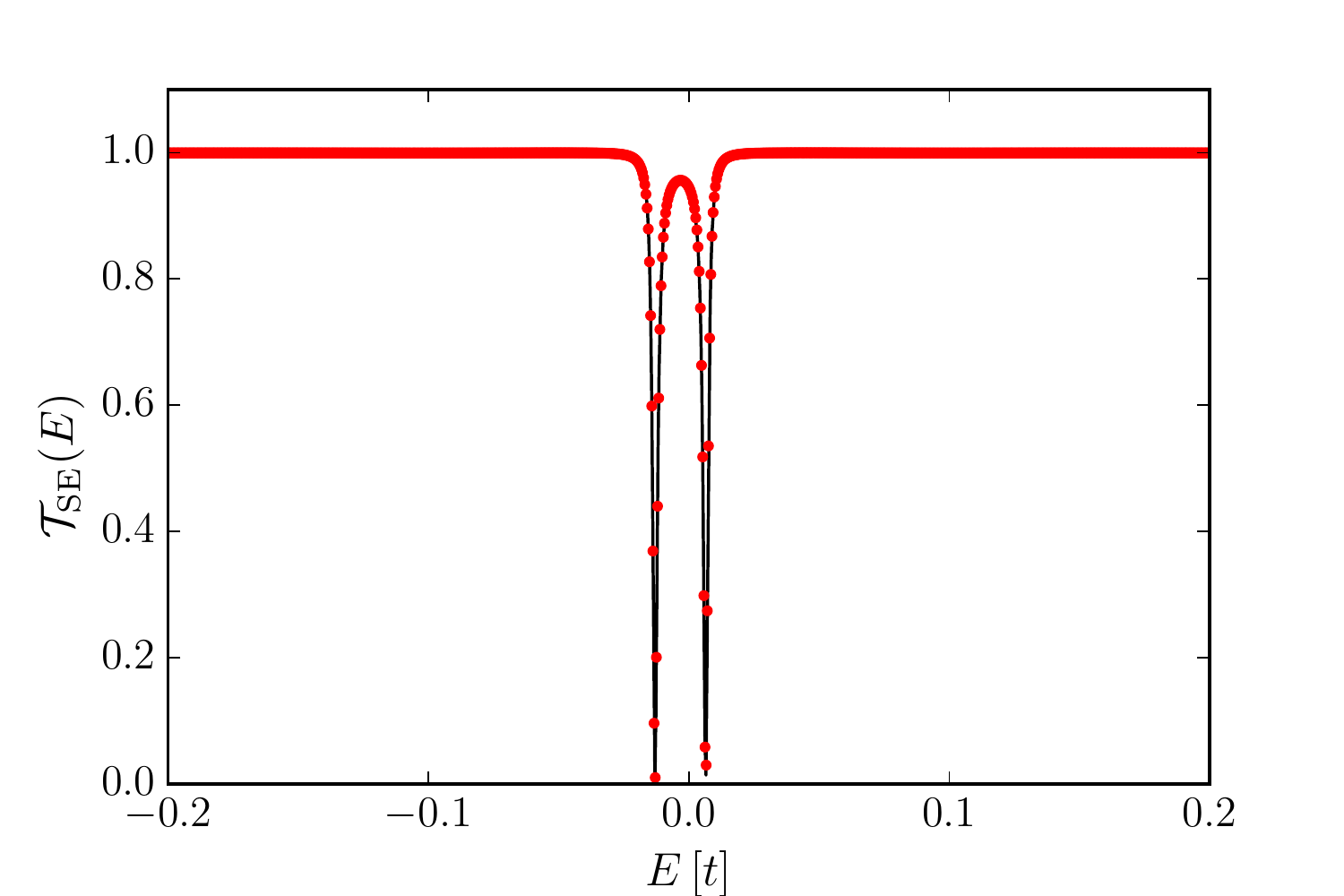}
\caption{(Color online) Numerical (red dots) and analytical (solid black line) results for the transmission ${\cal T}_{\rm SE}(E)$ relative to a single edge, as a function of energy $E$ (in units of $t$). Numerical results have been obtained for the same parameters relative to Fig.~\ref{fig:single_vacancy_transmission}, and $U/t=0.1$.\label{fig:single_vacancy_num_analytical_fit}}
\end{figure}

A least-square minimization procedure shows that Eq.~(\ref{eq:analytical_transmission}) fits very well our numerical data in the weak coupling regime $U/t \ll 1$. For example, Fig.~\ref{fig:single_vacancy_num_analytical_fit} shows a comparison between the numerical data ${\cal T}_{\rm SE}(E)={\cal T}(E)-1$ relative to a single edge for one vacancy and $U/t=0.1$, as reported in Fig.~\ref{fig:single_vacancy_transmission}, and Eq.~(\ref{eq:analytical_transmission}) with $\gamma= 0.0021$, $\Delta= 0.0196$, $E_{\rm a}= 0.0033$, and $\cos(\theta) = 0.0362$.
\section{Local density of states}
\label{sec:ldos}
We now discuss the behavior of the local density of states $D(E, {\bm r})$ at two different representative values of energy $E$.

\begin{figure}[t]
\centering
\includegraphics[width=0.5\textwidth]{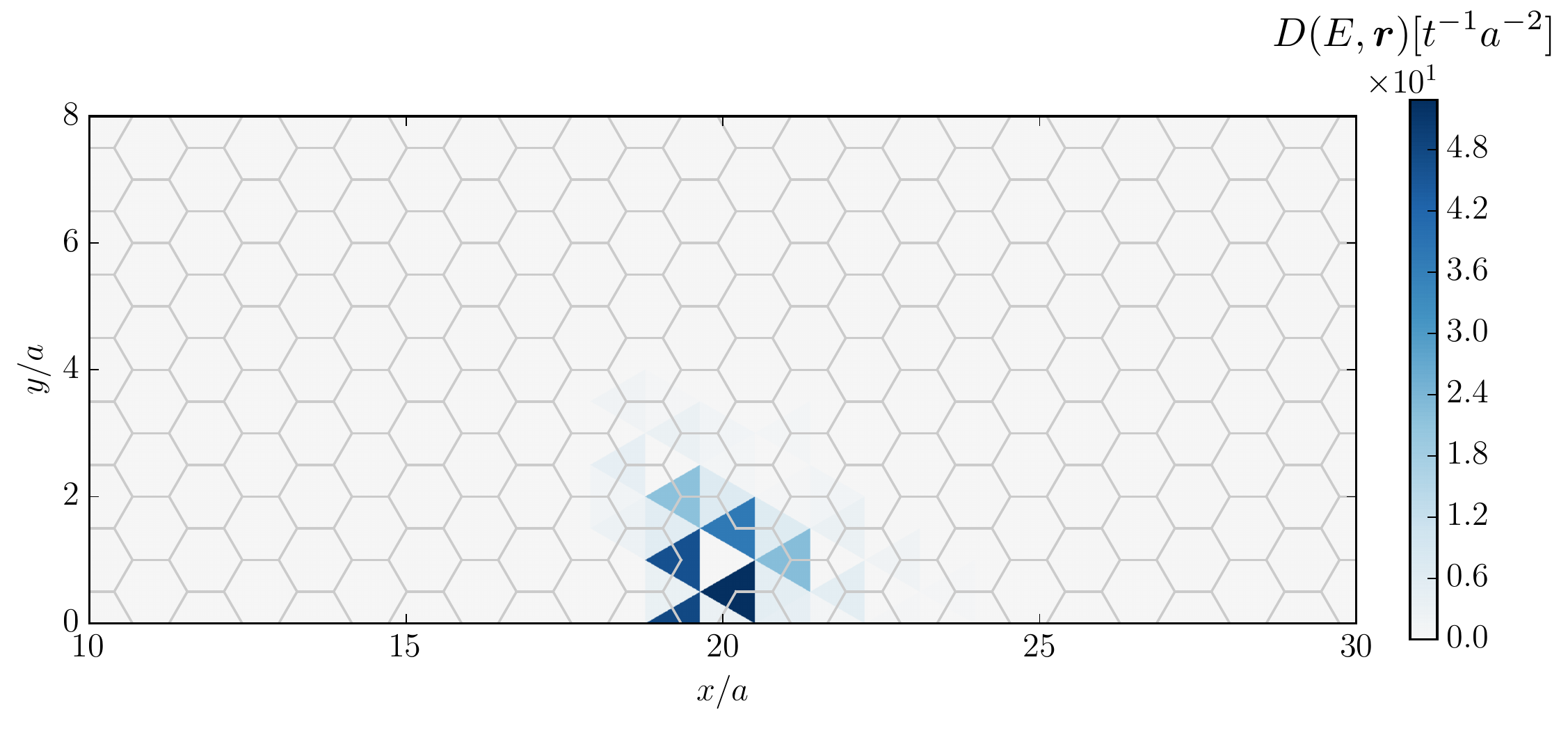} 
\caption{(Color online) Local density of states $D(E, {\bm r})$ (in units of $t^{-1} a^{-2}$) for a single vacancy, at the energy where the dip occurs in ${\cal T}(E)$ in Fig.~\ref{fig:single_vacancy_transmission} of the main text, i.e.~$E=-0.013t$. All parameters are the same as in Fig.~\ref{fig:single_vacancy_transmission} of the main text and $U/t = 0.1$. \label{subfig:1Vac_ldos_bound}}
\end{figure}

\begin{figure}[t]
\centering
\includegraphics[width=0.5\textwidth]{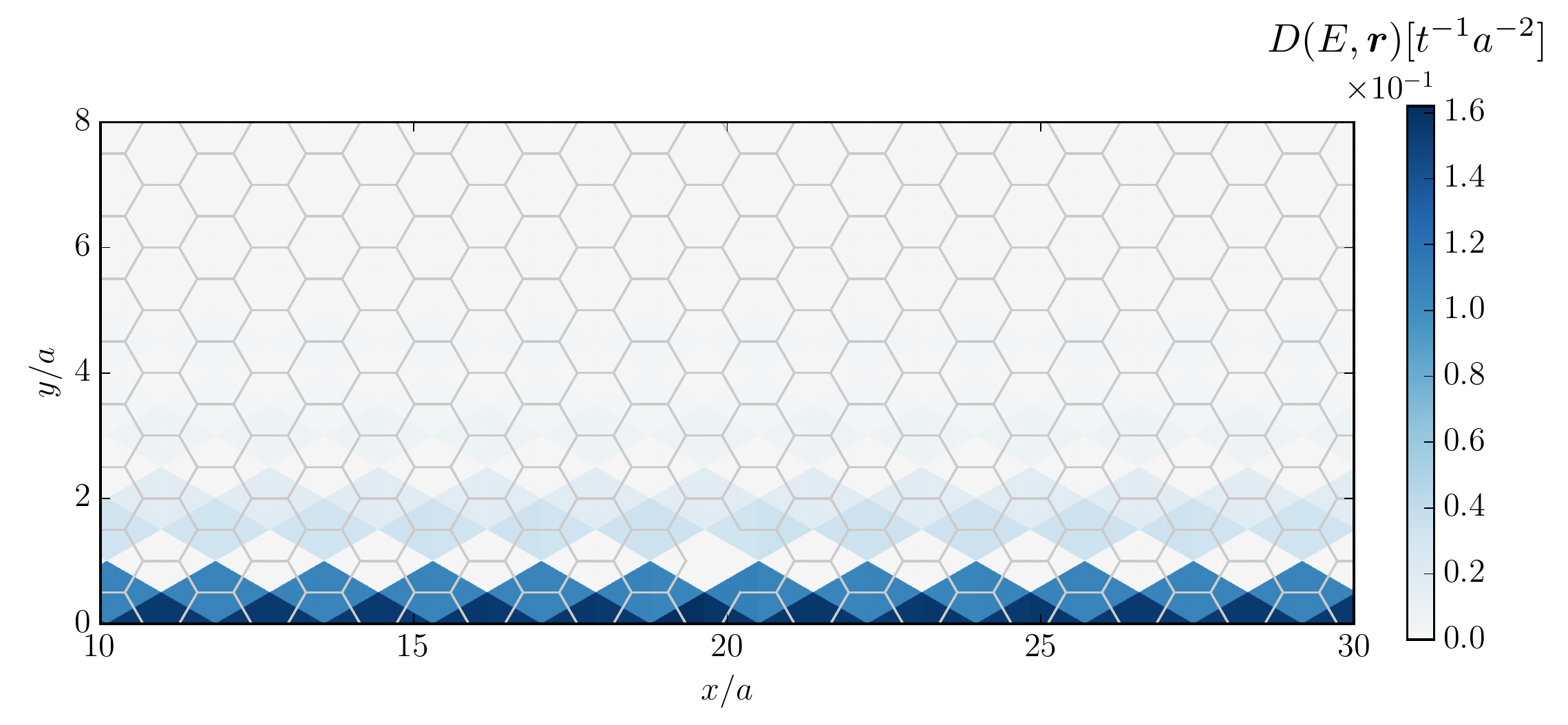} 
\caption{(Color online) Same as in Fig.~\ref{subfig:1Vac_ldos_bound} but for an energy $E=-0.2t$, i.e.~far from the one relative to the transmission dip. \label{subfig:1Vac_ldos_edge}}
\end{figure}

Fig.~\ref{subfig:1Vac_ldos_bound} shows $D(E, {\bm r})$ at $E=-0.013t$, which matches the energy of one of the dips for the case of a single vacancy and parameters as in Fig.~\ref{fig:single_vacancy_transmission} of the main text, with $U/t=0.1$. It is clear that $D(E, {\bm r})$ is localized around the position of the vacancy, where the local ground-state spin polarization is also finite (see Fig.~\ref{fig:magnetization}).
On the other hand, by choosing a value of energy far from the dip in the transmission, one finds that the corresponding states are delocalized along the edge (see Fig.~\ref{subfig:1Vac_ldos_edge}), and support transport.


\section{The impact of Rashba spin-orbit coupling}
\label{sec:rashba_soc}
\begin{figure}[t]
\centering
\includegraphics[width=0.5\textwidth]{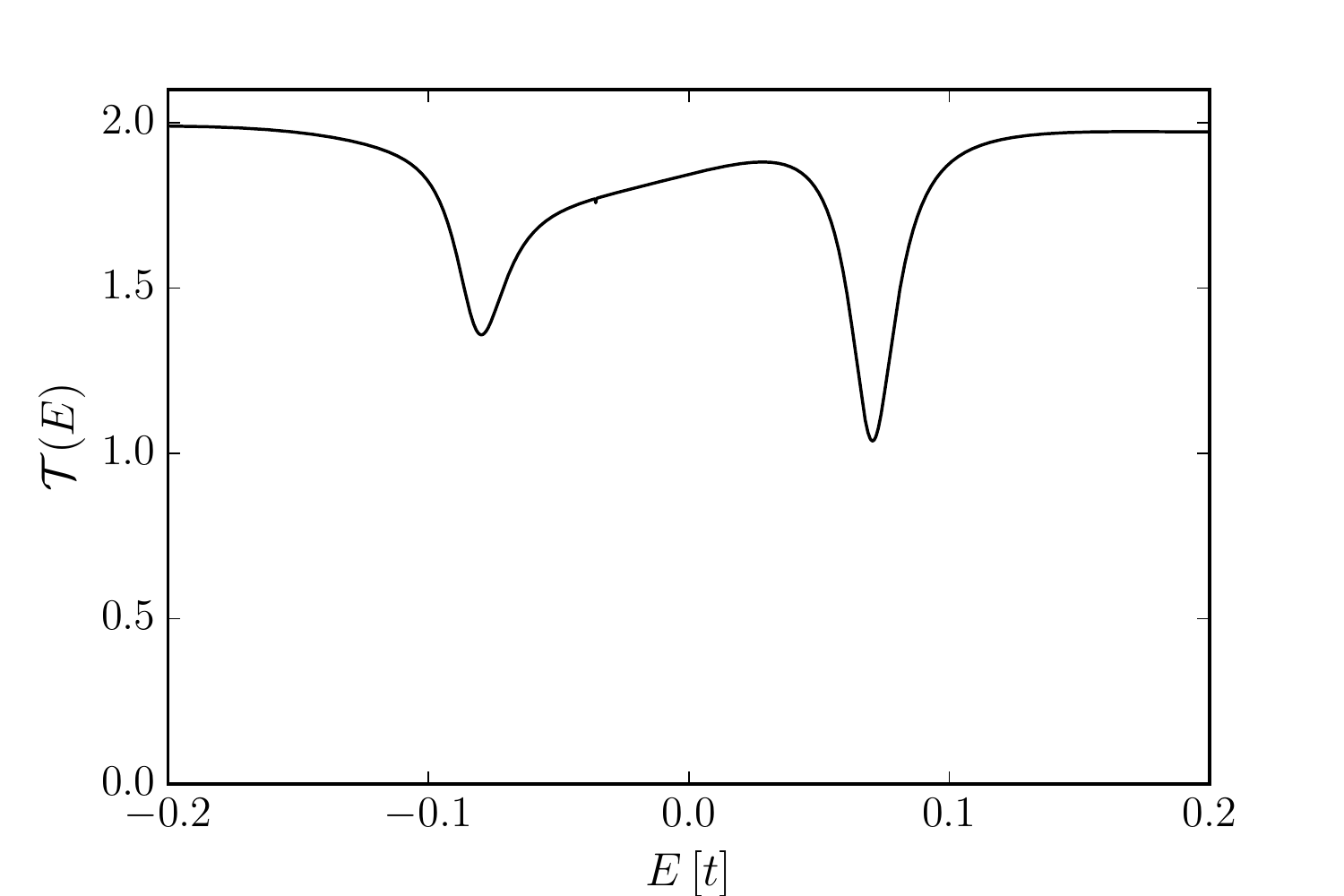}
\caption{Transmission ${\cal T}(E)$ as a function of energy $E$ (in units of $t$) for a single vacancy, 
obtained in presence of Rashba spin-orbit coupling. Numerical results in this figure have been obtained by setting 
$\lambda/t = 0.09$, $\lambda_{\rm R}/t = 0.15$, and $U/t = 0.5$. The position of the vacancy is as in the case of Fig.~\ref{fig:single_vacancy_transmission}.
\label{fig:Rashba}}
\end{figure}

The Kane-Mele Hamiltonian~\cite{kane_prl_2005_146802} can be enriched by a Rashba term of the form
\begin{equation}
{\cal H}_{\rm R} = i\lambda_{\rm{R}}\sum_{\langle i,j\rangle}c_{i}^{\dagger}\left( \boldsymbol{s}\times \boldsymbol{d}_{ij}\right)_{z} c_{j}~,
\end{equation}
which preserves the topological phase~\cite{kane_prl_2005_146802} for not too large values of $\lambda_{\rm R}$, with respect to $\lambda$.
Even in the presence of such term, a ground-state spin polarization still develops around a vacancy.
Fig.~\ref{fig:Rashba} shows the resulting transmission as a function of energy, calculated for $\lambda/t = 0.09$, $\lambda_{\rm{R}}/t = 0.15$, and $U/t = 0.5$.
Despite the large value of $\lambda_{\rm R}/t$, along with the dips broadening, the main difference with respect to the case where the Rashba term is absent (see Fig.~\ref{fig:single_vacancy_transmission}) is that now ${\cal T}(E)$ exhibits a sizable asymmetry with respect to $E=0$, as a consequence of the enhanced particle-hole asymmetry.
\end{document}